\documentclass[aps,pra,preprint,superscriptaddress,showpacs]{revtex4-1}
\usepackage[dvips]{graphicx}
\usepackage{datetime}
\def\be{\begin{equation}}
\def\ee{\end{equation}}
\def\e{{\rm e}}
\def\d{{\rm d}}
\def\({\left(}
\def\){\right)}
\begin{document}
\title{Spontaneous decay of small copper cluster anions Cu$_n^-$, $n=3-6$ 
on long time scales}
\author{K. Hansen}\thanks{klavshansen@tju.edu.cn}
\affiliation{Tianjin International Center of Nanoparticles and Nanosystems,
Tianjin University, 92 Weijin Road, Nankai district, Tianjin 300072, P.R.China}
\affiliation{Department of Physics and Astronomy, Aarhus University, 
8000 Aarhus, Denmark}
\affiliation{Department of Physics, University of Gothenburg, 41296 Gothenburg, Sweden}
\author{M.H. Stockett}
\affiliation{Department of Physics and Astronomy, Aarhus University, 8000 
Aarhus, Denmark}
\affiliation{Department of Physics, Stockholm University, 10691 Stockholm, Sweden}
\author{M. Kaminska}
\affiliation{Department of Physics, Stockholm University, 10691 Stockholm, Sweden}
\affiliation{Institute of Physics, Jan Kochanowski University, 25-369 
Kielce, Poland}
\author{R.F. Nascimento}
\affiliation{Department of Physics, Stockholm University, 10691 Stockholm, Sweden}
\affiliation{Centro Federal de Educacao Tecnologica Celso Suckow da Fonseca, 
Petropolis, 25620-003, RJ, Brazil}
\author{E.K. Anderson}
\affiliation{Department of Physics, Stockholm University, 10691 Stockholm, Sweden}
\author{M. Gatchell}
\affiliation{Department of Physics, Stockholm University, 10691 Stockholm, Sweden}
\author{K.C. Chartkunchand}
\affiliation{Department of Physics, Stockholm University, 10691 Stockholm, Sweden}
\author{G. Eklund}
\affiliation{Department of Physics, Stockholm University, 10691 Stockholm, Sweden}
\author{H. Zettergren}
\affiliation{Department of Physics, Stockholm University, 10691 Stockholm, Sweden}
\author{H.T. Schmidt}
\affiliation{Department of Physics, Stockholm University, 10691 Stockholm, Sweden}
\author{H. Cederquist}
\affiliation{Department of Physics, Stockholm University, 10691 Stockholm, Sweden}
\date{\today}\currenttime

\begin{abstract}
We have measured the spontaneous neutral particle emission from 
copper cluster anions (Cu$_n^-$, $n=3-6$) stored at cryogenic temperatures 
in one of the electrostatic
ion storage rings of the DESIREE (Double ElectroStatic Ion Ring ExpEriment)
facility at Stockholm University. The measured rate of emission 
from the stored Cu$_3^-$ ions follows a single power law decay for 
about 1 ms but then decreases much more rapidly with 
time. The latter behavior may be due to a decrease in the density of 
available final states in Cu$_3$ as the excitation energies of the decaying ions 
approach the electron detachment threshold. The 
emissions from Cu$_4^-$, Cu$_5^-$ and Cu$_6^-$ are 
well-described by sums of two power laws that are quenched by radiative 
cooling of the stored ions with characteristic times between a few and 
hundreds of milliseconds. We relate these two-component behaviors to 
populations of stored ions with higher and lower angular momenta.
In a separate experiment, we studied the laser-induced decay of Cu$_6^-$ 
ions that were excited by 1.13 eV or 1.45 eV photons after 46 milliseconds 
of storage. 
\end{abstract}
\maketitle

\section{Introduction}
The advent of a new generation of cryogenic electrostatic ion storage 
devices has greatly widened the feasible time range for studies of 
ion decay, opening the possibility to measure decay channels that 
are active on very long time scales 
\cite{Reinhed2009,Reinhed2010,Lange2010,thomas11,vonHahn2011,Nakano2012,schmidt13}. 
At DESIREE \cite{thomas11,schmidt13}, the 13 K operating temperature 
yields a residual gas 
pressure of roughly $10^{-14}$ mbar, allowing for the observation of 
processes with characteristic decay times on the order of thousands 
of seconds. This development puts special emphasis on radiative 
cooling processes, which are expected to be the dominant decay 
channels for molecules and clusters at long times. Fast radiative 
cooling of molecular ions stored in electrostatic rings was first 
observed in experiments on C$_{60}^-$ at the room temperature ring 
ELISA in Aarhus \cite{andersen96}. These fullerene ions were then 
found to cool with a rate exceeding those expected for vibrational 
cooling by two orders of magnitude and with characteristic 
cooling times of several milliseconds \cite{andersen96}.
Similarly, fast radiative cooling has been observed for anthracene 
cations (C$_{14}$H$_{10}^+$) \cite{martin13} stored in MINIring, 
a compact ion storage ring in Lyon \cite{bernard08}. Even faster 
radiative cooling rates, with corresponding timescales in the 
sub-millisecond range have been measured for small carbon cluster 
anions (C$_n^-$, $n=4,6$) at the TMU (Tokyo Metropolitan University) 
E-Ring \cite{ito14}. Also at TMU, slower cooling of C$_6$H$^-$ 
ions \cite{ito14} and of C$_5^-$ and C$_7^-$ clusters 
\cite{goto13,Najafian2014,Kono2015} on millisecond timescales was 
again observed. 

These previous experiments have revealed several unexpected 
features of the radiative cooling process. Of particular note is 
that it can proceed very effectively via electronic transitions, even 
for clusters as small as C$_{4}^{-}$ \cite{andersen96,martin13,Kono2015}.
It is the availability of low-lying electronic excitations in C$_4^-$
and C$_6^-$ and the lack of such excitations in C$_5^-$, C$_7^-$ and 
C$_6$H$^-$ that makes the cooling much more efficient in the former 
than in the latter cases. Recently, direct evidence for the emission 
of photons due to transitions from the lowest electronically excited 
optically active state in C$_6^-$ to the electronic ground state was reported 
\cite{Ebara2016}. Cooling of metal clusters is of special interest 
because in addition to evaporative cooling through emission of 
electrons, atoms, or molecules, metal clusters can radiate and thus 
cool via any of the following radiative mechanisms: Decay from a single 
particle-hole electronic excitation; Excitation of a short-lived 
surface plasmon resonance; or by vibrational and rotational cooling.
The first two are unique to clusters of metals and other systems 
with delocalized electrons, including carbon and carbon-based 
molecules. Studies of metal clusters are thus the key to understand 
the importance of these types of radiative cooling processes.  

Only a few radiative cooling experiments have so far been carried 
out using {\it cryogenic} electrostatic ion storage devices. These are 
experiments on Al$_n^-$, $n=4-5$ clusters \cite{froese11} and on 
SF$_6^-$ \cite{menk14} at the Cryogenic Trap for Fast ion beams, 
CTF, in Heidelberg, and a recent study of Cu$_n^-$, $n=4-7$, 
also in the CTF \cite{BreitenfeldtPRA2016}. In the 
present paper we report measurements of the spontaneous emission rate 
of neutral particles
from the small copper cluster anions Cu$_n^-$, $n=3-6$, stored in one 
of the DESIREE ion-storage rings at Stockholm University 
\cite{thomas11,schmidt13}. These measurements carry information on 
the rates of radiative cooling processes on different time scales. 

\section{Experimental}

The cluster anions were produced in a cesium-sputter ion source using a solid 
copper cathode and accelerated to 10 keV. A bending magnet was used to 
select the desired mass before injection into the ring. For the smallest 
cluster sizes, the smallest mass isotopologues were used. No sign 
of hydrogenation was found and for the larger clusters the most intense 
peak in the isotope distribution was used.
\begin{figure}
\centering
\includegraphics[width=8cm,angle=0]{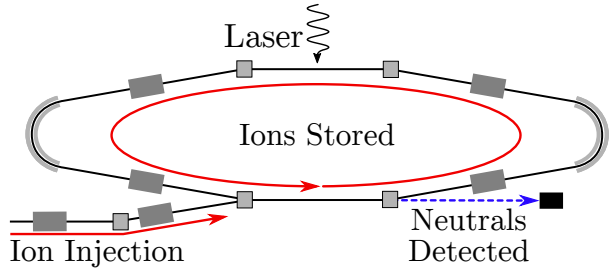}
\caption{One of the DESIREE storage rings with indicated charged and 
neutral particle trajectories. 
The ions were injected into the ring using a set of deflector plates. 
Before the ion bunch has made one turn, the voltage on the 
deflector is switched to a voltage such that the ion beam is stored in 
the ring. Neutrals from spontaneously decaying, excited Cu$_n^-$ 
clusters with $n=3-6$ or laser-excited Cu$_6^-$ clusters were detected 
by a multi channel plate detector at the end of one of the straight 
sections of the ring. The crossed laser beam, used for experiments with 
Cu$_6^-$, is indicated.
\label{fig_desiree}}
\end{figure}
Ions were injected, stored, and dumped every 0.1-10 s depending on the 
time scale of interest. The intrinsic ion storage 
lifetime in DESIREE is much longer than this, as 
demonstrated by measurements of metastable excited states in 
atomic anions where a storage lifetime of 30 minutes was measured 
for a 10 keV kinetic energy Te$^-$ beam \cite{BackstromPRL2015}. 
For the Cu$_n^-$, $n=3-6$, cluster anions studied here, the signal 
due to spontaneous decays of excited ions always becomes negligibly 
small within a few seconds after injection and time windows of the 
order of ten seconds are thus sufficient in these particular cases. 

In the experiment, we detect the neutral particles that are produced 
through electron detachment and/or unimolecular dissociation in 
spontaneous decay processes or, in separate experiments, from decays 
induced by absorption of a photon from a laser pulse.
In both types of experiments, neutral particle counts were 
recorded as functions of time, $t$, with the detector placed along the 
line of sight of one of the straight sections of the ion-beam storage 
ring as shown in Fig. \ref{fig_desiree}. Since the detector only 
registers neutral particles, it is not possible to assign the decay 
channel from the signal. Identification of specific decay channels is, 
however, not necessary for the determination of cooling times. The 
decay curve for Cu$_3^-$ was recorded using time windows of 200 ms and 
10 seconds. After normalization, the decay curves agree with each other 
in the region where they overlap. The spectra for Cu$_{4,5,6}^-$ 
were recorded with time windows of a few seconds.

In addition to these measurements, crossed-beam laser excitation 
experiments were performed 
on Cu$_6^-$ with the wavelengths 850 nm (1.45 eV) and 1100 nm (1.13 eV), 
as shown in Figure \ref{fig_desiree}. The delayed 
photo-induced signals for 
the smaller clusters were found to be significantly weaker than the 
ones for Cu$_6^-$ and are not reported here. For both wavelengths the 
laser was fired 46 ms after injection of the Cu$_6^-$ beam. 
At this time the spontaneous 
decay from hot clusters produced in the source has almost disappeared 
and the measurement is close to background-free.

\section{Results}
\subsection{Overview}
\begin{figure}
\centering
\includegraphics[width=0.45\textwidth]{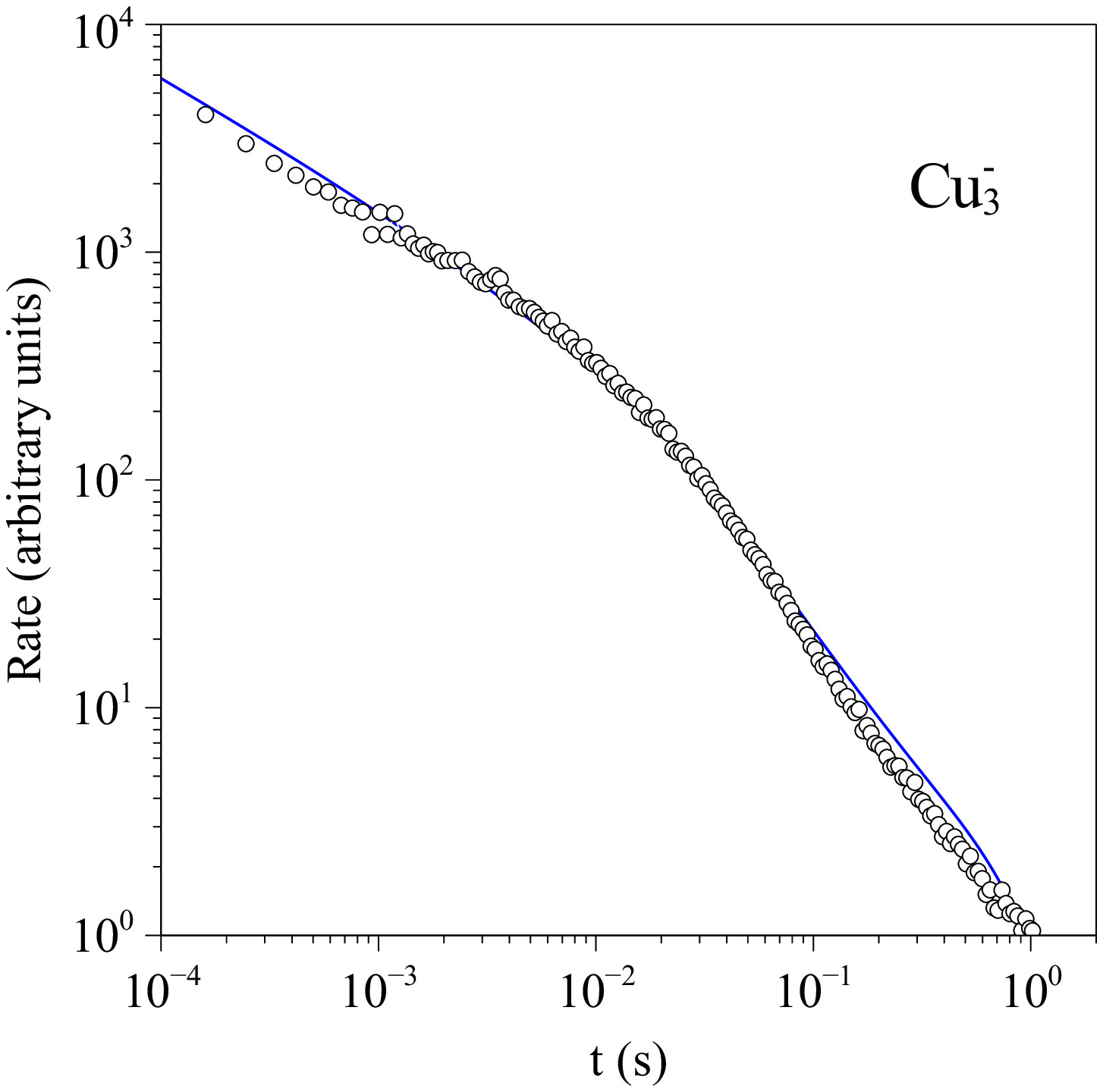}
\caption{Rates of neutral particles leaving a stored beam of 
Cu$_3^-$ ions as functions of the time, $t$, after their production 
in the ion source. The points are measured values and the blue line is the 
calculated decay rate $R(t)$ multiplied by $t^{0.67}$ as discussed 
in the main text.
\label{fig_cu2-3}}
\end{figure}

Decay curves of the stored ions are shown in Figures \ref{fig_cu2-3}, 
\ref{fig_cu4}, \ref{fig_cu5} and \ref{fig_cu6}, in double logarithmic 
plots of the measured neutral emission rate for the stored ions 
vs. the time $t$ after production in the source. 
The data (recorded number of events)  are binned in time intervals 
that increase linearly in width with $t$. These binned data are then 
divided by the width of the time bin to obtain the rate. This gives an 
even distribution of rate values along the horizontal axes in 
Figures \ref{fig_cu2-3}, \ref{fig_cu4}, \ref{fig_cu5}, and \ref{fig_cu6}.
The background is due to detector 
dark counts and collisions with residual gas. The largest component 
of this background is due to the detector dark counts, which has 
been measured separately, and which is subtracted prior to the 
binning of the data.

\subsection{Cu$_3^-$}
The Cu$_3^-$ spectrum shown in Fig. \ref{fig_cu2-3} is 
well-described by a power law at short times. At longer times 
the curve has a much steeper slope. This apparent two-component 
power law decay could in principle be accounted for by 
the small number of vibrational degrees of freedom, as indicated 
by a direct numerical calculation. The general expression for 
the radiation-free spontaneous particle emission rate, 
$R(t)$, is
\be
R(t) \propto \int_0^{\infty} g(E)k(E){\e}^{-k(E)t}{\d}E,
\label{eq_rateint}
\ee
in which $k(E)$ is taken to be the electron emission rate constant for ions with internal 
excitation energy $E$, and $g(E)$ is the distribution of excitation 
energies. When this distribution is sufficiently broad, {\it i.e.} varies 
much slower with $E$ than $k(E)$, Eq.\ref{eq_rateint} 
can be approximated by a power-law decay rate \cite{hansen01},
\be
R(t) \propto t^{p},
\label{powerlaw}
\ee
with $p \approx -1$. In general, this approximate result holds irrespective 
of whether the process is the statistical emission of an electron, an atom, or 
a molecular fragment.
It may fail, however, for clusters with small heat 
capacities, such as Cu$_3^-$ (cf. the discussion in section \ref{subsec-ple}
below). A quantitative estimate of the deviation 
from the power law in Eq. \ref{powerlaw} requires the calculation of an explicit expression 
for the rate constant, $k(E)$, that enters into Eq.\ref{eq_rateint}. 
Here, the calculation 
was done with the detailed balance rate constant for electron 
emission \cite{Weisskopf,book}:
\be
\label{rateconstant}
k(E,\varepsilon) \d\varepsilon =
\frac{2m_e}{\pi^2 \hbar^3} \varepsilon \sigma_c
\frac{\rho^{(0)}(E-E_a-\varepsilon)}{\rho^{(-)}(E)} \d \varepsilon,
\ee
in which $m_{e}$ is the electron mass, $\rho^{(0)}$ and $\rho^{(-)}$ are 
the level densities of the neutral and the anion, $E$ is the Cu$_3^-$ excitation 
energy, $E_a$ the electron affinity of Cu$_3$, 
$\sigma_{c}$ is the Cu$_{3}$ electron capture cross-section, and $\varepsilon$ 
is the kinetic energy of the decay channel - effectively the kinetic 
energy of the electron.
The factor of two in Eq. \ref{rateconstant} accounts for the spin degeneracy 
of the emitted electron.
The level densities were calculated with the Beyer-Swinehart algorithm 
\cite{BeyerSwinehart} using the following four vibrational frequencies 
determined by density functional theory (B3LYP/LANL2DZ): 
230 cm$^{-1}$, 142 cm$^{-1}$, 
and 52 cm$^{-1}$ (doubly-degenerate). 
The same frequencies were used for both Cu$_3$ and Cu$_3^-$. For the 
electron affinity, which acted as the activation energy, the 
experimental value 2.45 eV \cite{Leopold1987} was used. The electron 
attachment cross-section, $\sigma_c$, was set to the constant value of 
$1.5\cdot 10^{-21}$ m$^2$. This small cross section is strongly reduced 
from the Langevin value as an assumed effect of a 
small electron-cluster sticking coefficient. 
This assumption is made here as the density of states in Cu$_3^-$ 
may well be very low due to a very narrow range of electron energies.
In the 
calculation of the total rate constant, $\varepsilon$ is integrated 
out by a numerical summation over the states;
\be
\label{KERint}
k(E) = \sum_i k(E,\varepsilon_i) \delta \varepsilon,
\ee
in which the sum runs over the discretized values of $\varepsilon$ 
and $\delta \varepsilon$ is the energy resolution of the level 
density defined by the Beyer-Swinehart calculation. In addition to 
the rate calculated by numerical integration of Eq.\ref{eq_rateint} 
with the rate constant from Eq. \ref{rateconstant} and \ref{KERint}, 
the calculated function, $R(t)$, is multiplied by the 
time to the power 0.67 in Fig. \ref{fig_cu2-3}. This power is a fit 
factor which could possibly be related to the shape of the 
excitation energy and angular momentum distributions
of the clusters produced in the source. 

We assign the deviation from a single power law decay 
behavior of Cu$_3^-$ to the low density of vibrational states in 
neutral Cu$_3$ close to the vibrational ground state. The cross-over 
from one slope to the other could be a result of the freezing 
out of a single vibrational degree of freedom. However, both slopes
will be too steep without the fit factor $t^{0.67}$. Furthermore, 
the second part of the curve requires the electrons to be captured 
in the inverse process with the reduced Langevin cross section 
mentioned above. The effect of the reduction is to postpone the 
freezing out of the last vibrational degree of freedom and instead 
make the next-to-last freezing out observable in the experimental 
time window. Apart from this last shift, the mechanism proposed is 
similar to the one
made in connection with the CTF study of SF$_6^-$ in Heidelberg 
\cite{menk14}. The main difference to those data is that not all 
final state vibrations freeze out at the same excitation 
energy in the copper trimer. Therefore the decay may continue to 
follow a power law but with a different slope than at earlier times. 

Finally, we note a very slight change in the slope for the 
experimental data at times longer than 0.1 seconds in Fig. \ref{fig_cu2-3}.
A weak component from a second distribution of internal energies 
of Cu$_3^-$ that could be due to trimer ions with different 
conformations and/or angular momentum distributions cannot be ruled out.

\subsection{Cu$_{4,5,6}^-$}
\begin{figure}[h]
\centering
\includegraphics[width=0.45\textwidth]{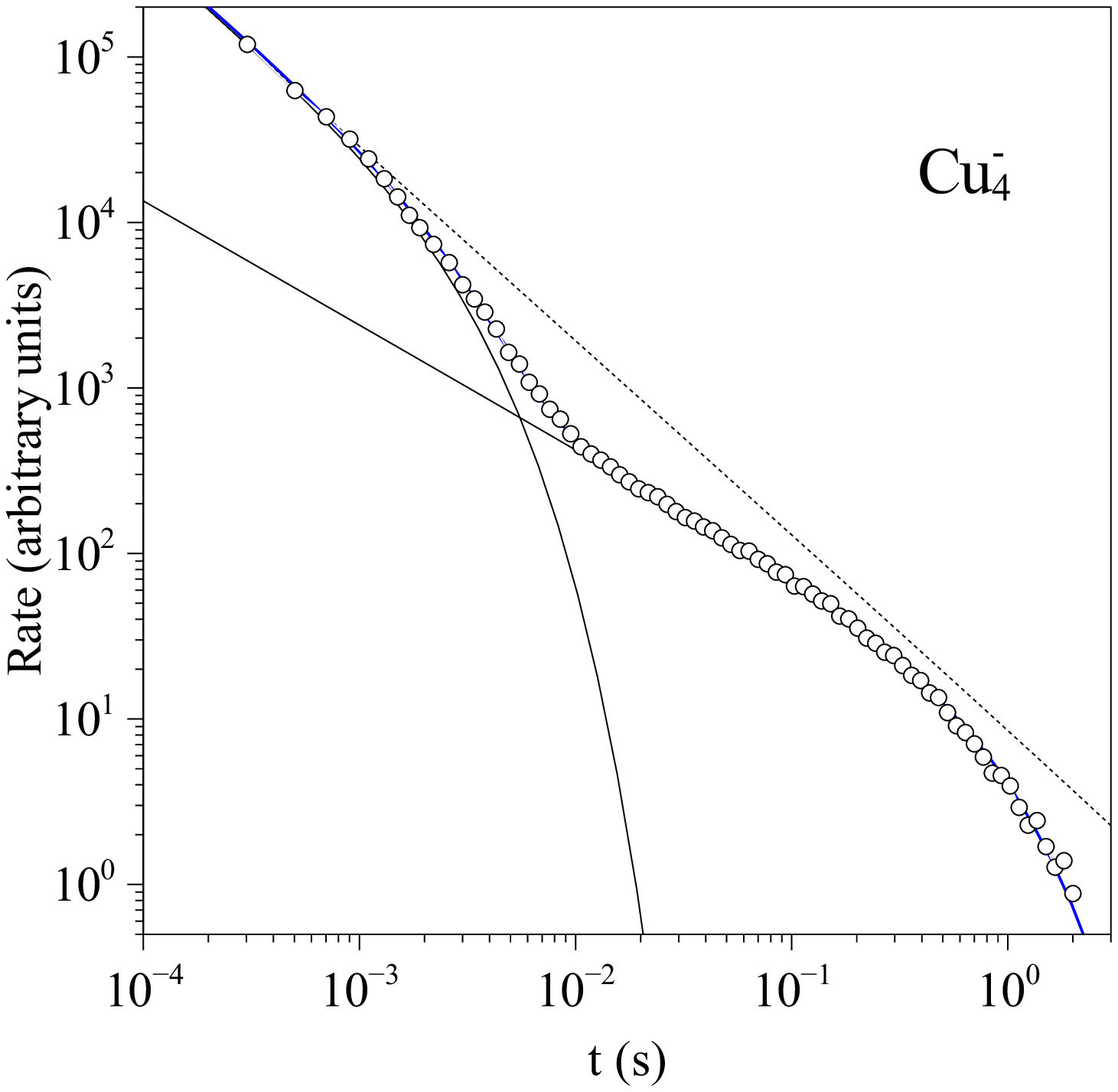}
\caption{Rates of neutral particles leaving a stored beam of Cu$_4^-$.
Two quenched power-law decays are needed to fit the data. 
The blue line is the fit with Eq.\ref{twocomp}, and the thin 
black lines the individual contributions from the two terms 
in the same equation. The shortest times are excluded from 
the analysis for both components because the detector is 
saturated during the first few turns in the ring. The dotted 
line is the simulated decay rate without radiative cooling 
for a single population, calculated analogously to the line 
in the Cu$_3^-$ frame. The parameters for the fits are given 
in Table \ref{tab_fitparams}.
\label{fig_cu4}}
\end{figure}

\begin{figure}[h]
\centering
\includegraphics[width=0.45\textwidth]{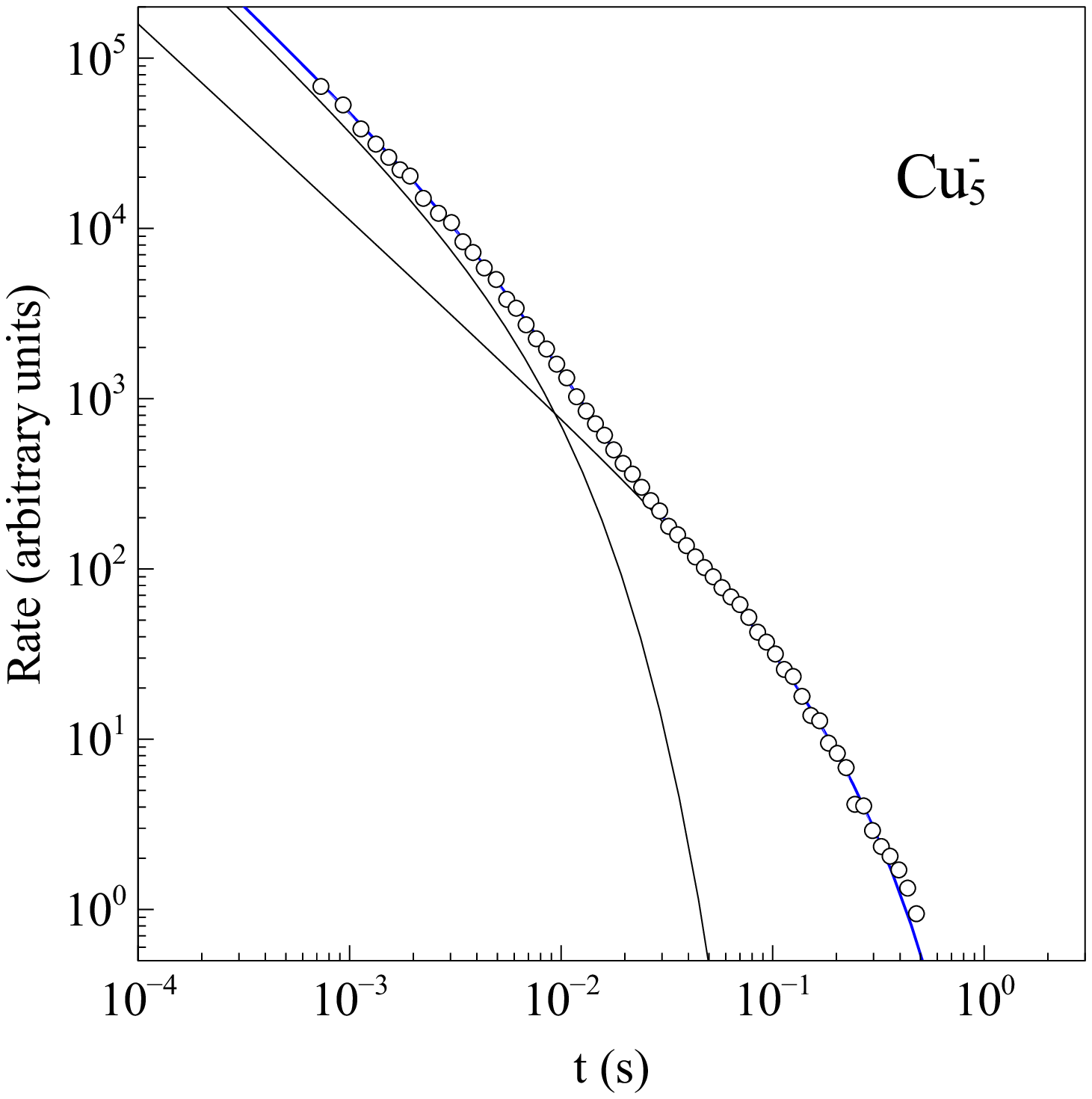}
\caption{Rates of neutral particles leaving a stored beam of Cu$_5^-$ ions.
The lines are analogous to the ones for Fig. \ref{fig_cu4}. 
The parameters for the fits are given in Table \ref{tab_fitparams}.
\label{fig_cu5}}
\end{figure}

\begin{figure}[h]
\centering
\includegraphics[width=0.45\textwidth]{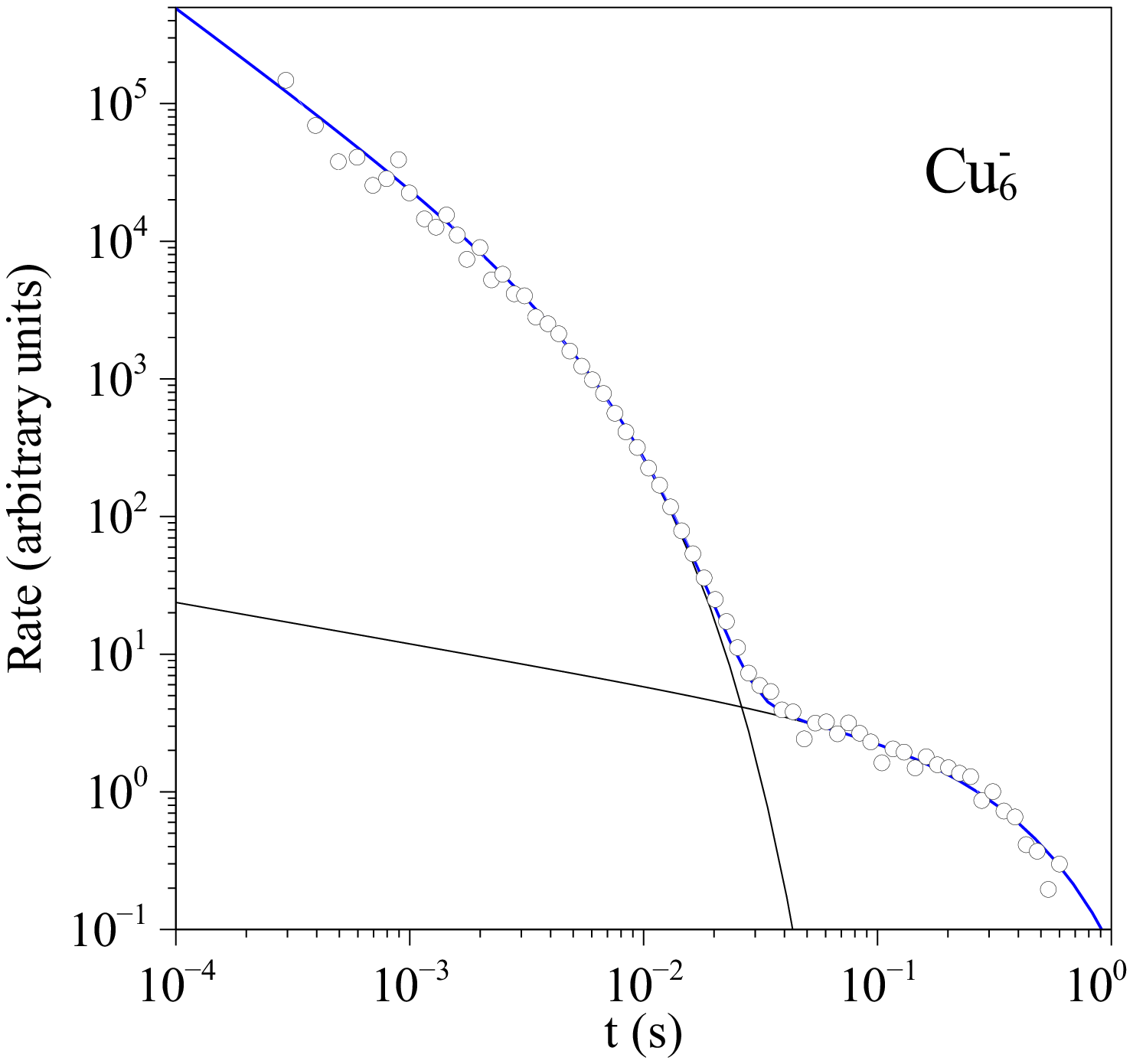}
\caption{Rates of neutral particles leaving a stored beam of Cu$_6^-$ 
ions. The lines are analogous to the ones for Fig. \ref{fig_cu4}. 
The parameters for the fits are given in Table \ref{tab_fitparams}.
\label{fig_cu6}}
\end{figure}

The measured neutral particle emission signals for Cu$_4^-$,
Cu$_5^-$ and Cu$_6^-$ are more complex than that of Cu$_3^-$ 
and for Cu$_4^-$, the measurements span more than four orders 
of magnitude in time (see Figure \ref{fig_cu4}). For Cu$_5^-$ and Cu$_6^-$ 
the range is nearly four orders of magnitude in time (see Figs. 
\ref{fig_cu5} and \ref{fig_cu6}.

A calculation similar to the one for Cu$_3^-$ shows a qualitatively 
similar effect for Cu$_4^-$, albeit so strongly reduced that the decay 
rate is very close to a single power-law (cf. the dotted line in 
Fig. \ref{fig_cu4}), and therefore it cannot explain the experimental 
data. For Cu$_{4}^{-}$, Cu$_{5}^{-}$ and Cu$_6^-$, the data are instead 
well-represented by a sum of two curves 
\be
R(t) = a_1t^{-1+\delta_1}e^{-t/\tau_1} + a_2t^{-1+\delta_2}e^{-t/\tau_2}.
\label{twocomp}
\ee
As mentioned above, the power law decay at short 
times is an effect of the broad internal energy distribution 
of the cluster ions from the source \cite{hansen01}, and the 
exponential decay the consequence of radiative cooling. 

\section{Discussion}
\subsection{Role of Angular Momentum}
The existence of two components of the decay curves for 
Cu$_{4,5,6}^{-}$ will be discussed in terms of two distinct 
populations of ions, characterized by different angular momenta.
The observation of two beam components with very different 
lifetimes will then be linked to a conserved quantity that effectively 
preserves the integrity of the two populations, and endows them 
with distinct and conserved properties. Of primary importance
for the following reasoning is the different lowest energy 
geometries of species with different angular momenta. The sputter 
source that we use is of a type known to produce dimer- and larger anions 
in high rotational states
\cite{Fedor05,froese11,AvivPRA2011,menk14,KaflePRA2015}, 
and this is most likely the case also in the present experiment. 
We therefore tentatively assign the two characteristic times for 
each one of the three cluster sizes $n=4,5,6$ to species with higher 
and lower angular momentum.

\begin{figure}[h]
\centering
\includegraphics[width=0.45\textwidth]{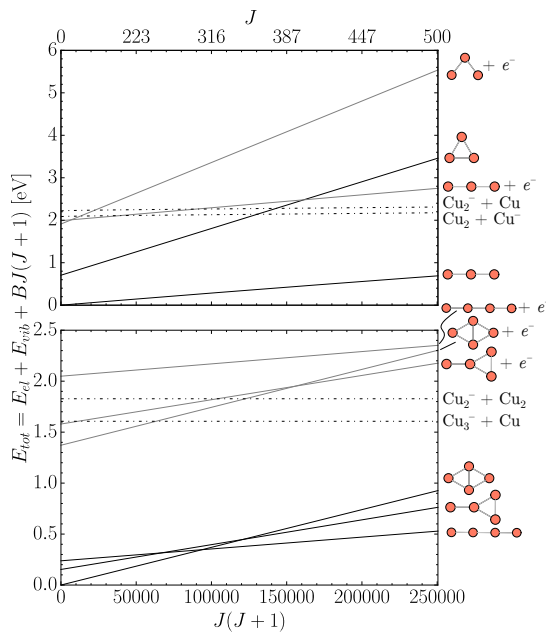}
\caption{Calculated relative excitation energies (black lines) 
for two Cu$_3^-$ (upper panel) and three Cu$_4^-$ (lower panel) 
conformers in their lowest vibrational states as functions of $J(J+1)$
where $J$ is the rotational angular momentum quantum number. The 
corresponding electron detachment limits (full gray lines) and the 
lowest-energy fragmentation channels (dash-dotted lines) are also 
indicated. There are progressions of vibrational states (not indicated) 
from the lowest state (black lines) of each conformation, reaching 
into the continuum beyond the corresponding detachment limit. 
No states exist below the lowest black line in each plot (the 
yrast line). Ions in excited states that are below both the 
lowest-energy detachment limit and the lowest dissociation limit 
are bound and cannot produce neutrals. 
The rotational barrier heights for the fragmentation channels
increase with $J$ and the dash-dotted lines should in principle have 
small non-zero slopes. We have calculated this effect for the 
dissociating Cu$_3^-$ clusters, but the effect is not visible on the 
scale shown here (the same is expected to be true for Cu$_4^-$).
\label{skematisk}}
\end{figure}

The suggested mechanisms are discussed with the aid of 
Fig. \ref{skematisk}. 
In this figure we show relative excitation energies for Cu$_3^-$ 
and Cu$_4^-$ ions in different conformations and as functions of 
their rotational angular momenta, $J$, with cluster structures 
from Density Functional Theory calculations (B3LYP/LANL2DZ). These 
energies are given relative to the energies of the most stable 
non-rotating isomers of Cu$_3^-$ and Cu$_4^-$ in their vibrational 
ground states. For each anion conformation there are progressions 
of vibrational states from the yrast line and upward. Conformers 
can convert into each other as long as energy and angular momentum 
are conserved, but the higher energy conformer populations 
are suppressed by their small level densities relative to those of 
the ground states.

Copper trimer anions in states with total excitation energies and 
$J$-values above the lowest gray line in the upper panel of 
Fig. \ref{skematisk} may decay through electron detachment or through 
fragmentation. The latter process is, however, most likely slowed down 
as it leaves the product with only a single vibrational degree of freedom 
while electron emission gives products with four vibrational degrees of 
freedom. Electron detachment will thus be entropically favored in this 
upper region of energies and $J$-values. Trimer 
anions with total energies below the same gray line, but above the 
dot-dashed lines, may only decay through dissociation channels.

For the Cu$_4^-$ ions, the situation is more complicated. There are three 
conformer detachment limits that are close in energy and mostly lie above 
the fragmentation limits as indicated in the lower panel in 
Fig. \ref{skematisk}.
The three conformers have linear, rhombic, and Y-shaped forms with the 
detachment limit for the linear one being highest in energy for all 
values of $J$ below 500. The detachment limits for the rhombic and Y-shaped 
forms cross near $J = 400$ – the latter has the lowest energy for $J > 400$ 
while the rhombic form is lower in energy for $J < 400$. There is a region 
of total and rotational energies where $J < 300$ and where the detachment 
limit for the rhombic form lies below the lowest fragmentation limit.  
Only electron detachment processes are possible in this ($J < 300$) region.

These relations between conformer detachment limits for Cu$_3^-$ 
and Cu$_4^-$ indicate that the decay behaviors can be different for 
Cu$_4^-$ ions in low and high rotational states while such a situation 
is less likely to occur for Cu$_3^-$. In this 
scenario, it is possible that the two distributions that appear in the 
data for Cu$_4^-$ (see Fig. \ref{fig_cu4}) are related to detachment from 
Y-shaped anions above $J=400$ and detachment from rhombic anions for 
$J < 400$, respectively. However, when we also consider the fragmentation 
channels - assuming for the moment that the corresponding rates 
are not negligible in 
relation to the present experimental time scales - we note that fragmentation 
is the lowest-energy channel leading to neutral products for $J > 300$ 
while the detachment limit is lowest in energy for $J < 300$ in Cu$_4^-$. 
The situation is different for Cu$_3^-$ where fragmentation has the 
lowest excitation energy (or close to) over a much wider range of $J$-values. 
In any case, it appears that different parts of the distribution, belonging 
to ions in high and low $J$-states may give rise to the double structures 
that we measure in the decay of Cu$_4^-$. 

Summarizing the suggested explanation, we have seen that the 
trimer decay can be explained without invoking two populations, 
consistent with the linear ground state structure seen in quantum 
chemical calculations including the present ones. The larger 
clusters with theoretically 
more compact ground states, on the other hand, display
double decay curves, consistent 
with anionic clusters belonging to two populations with
different angular momenta and conformations - linear 
species at high $J$ and more compact (rhombic) species at low $J$.
At the high angular momenta the neutral Cu$_4^-$ clusters may be rhombic or 
Y-shaped, depending on the precise value of $J$.
These trends for Cu$_4^-$ are likely to apply to the two larger 
clusters, also, as these are not expected to be linear for low $J$-values.

\subsection{Radiative Cooling}
We now turn to the quasi-exponential decrease of the Cu$_4^-$, Cu$_5^-$, 
and Cu$_6^-$ decay curves after a few ms, ascribed to radiative cooling. 
Radiative cooling, {\it i.e.} photon emission from excited vibrational 
\cite{andersen03}, electronic \cite{martin13}, or plasmonic \cite{andersen96} 
states, depletes the population of hot ions without producing 
neutral particles and effectively quenches the power law decay rate.
Depending on the magnitude of the energies of the emitted photons, 
the resulting spontaneous decay rate may either vary as \cite{hansen01}
\be
\label{power-rad2}
R(t) \propto \frac{t^{\delta}}{{\e}^{t/\tau}-1},
\ee
which is valid for small photon energies, or as 
\cite{Kono2015}
\be
\label{power-rad}
R(t) \propto t^{-1+\delta} {\e}^{-t/\tau},
\ee
for larger photon energies. Here, "large" photon energies are 
those for which the emission of a single photon suppresses any 
further unimolecular decay.
For short times, $t\ll \tau$, both expressions reduce to the power 
law, $R \propto t^{p}$, with $p=-1+\delta$, and for long times to a 
quasi-exponential decay. 

The photon energy required to quench the decay can be compared 
with the vibrational energy quantum. The (room temperature) 
Debye temperature of 
bulk copper is 310 K\cite{DebyeT}, corresponding to a quantum energy 
of 0.027 eV. 
Even for emission of such a comparatively low energy photon,
the effect for the present clusters will be a quenching of further 
unimolecular decay, corresponding to a decay rate following 
Eq. \ref{power-rad} \cite{Najafian2014}. For electronic transitions, 
the quenching effect is even stronger because of the much larger 
energies of the emitted photons. In either case, we can use 
Eq.\ref{power-rad} for the fits of the data reported here. The 
fitted effective quenching times, $\tau$, then directly give the 
photon emission rate constants without any further analysis. 

The fitted radiative time constants in table \ref{tab_fitparams} 
are mean values of the thermally populated, vibrationally excited 
states. The detailed balance expression for the photon emission 
rate constant, $k_p$, can be represented as \cite{Andersen2002,book}
\be
\label{photonrate}
k_p (E) = \int \frac{8\pi\nu^2}{c^2} \sigma_{abs}(\nu)
\frac{\frac{\rho(E-h\nu)}{\rho(E)}}
{1-\frac{\rho(E-2h\nu)}{\rho(E-h\nu)}} d\nu,
\ee
in which $\nu$ is the cyclic frequency of the emitted photon and 
$c$ the speed of light. We have assumed an excitation energy independent 
absorption cross-section, $\sigma_{abs}(\nu)$. Finally, $\rho$ is the 
level density, calculated with all modes except the emitting states.
From the expression it is clear that thermal photon emission is an activated 
process. The broad energy distribution and resulting spread 
in the radiative rate constant may therefore potentially render 
the radiative decay non-exponential, in analogy to the non-radiative 
electron emission. The effect will, however, not appear here 
because the radiative rate constant, $k_p$, is much less energy dependent 
than the unimolecular decay constant, and a description in terms 
of an energy-independent photon emission rate constant is a 
very good approximation. For an illustration of this point, see 
Fig.3 of \cite{Kono2015}.

The shorter of the characteristic cooling times for Cu$_4^-$ (2.6 ms),
for Cu$_5^-$ (7.5 ms) and for Cu$_6^-$ (5.9 ms) are all rather close
to each other.
These lifetimes, as well as their associated power law exponents 
$p=-1+\delta$ discussed below, are very similar to those 
measured for Al$_4^-$ and Al$_5^-$ \cite{froese11}.
These ions were also produced with a sputter source and stored in a 
cryogenic ion beam trap. Time constants of several 
milliseconds are expected for vibrational transitions, which suggests 
emission of infrared photons as the source of this cooling.
The assignment of the two different time constants requires 
a more detailed analysis which is outside the scope of this paper.

\subsection{Power Law Exponents} 
\label{subsec-ple}
Turning finally to the question of the initial power law decay, 
we address the deviations of $p$ from minus unity. The fitted values 
for the spontaneous decay of the three largest clusters are summarized 
in Table \ref{tab_fitparams}.
\begin{table}
\begin{tabular}{|c|cc|cc|c|}
\hline
& \multicolumn{2}{c|}{Short} & \multicolumn{2}{c|}{Long} & \multicolumn{1}{c|}{}\\
$N$ & $\delta$ & $\tau$ (ms)& $\delta$ & $\tau$ (ms) & VDE (eV)\\
\hline
4 & -0.1 & 2.6 & 0.25 & 830 & 1.45\\
5 & -0.2 & 7.5 & -0.15 & 180 & 1.94\\
6 & -0.28 & 5.9 & 0.7 & 330  & 1.96\\
\hline
\end{tabular}
\caption{Parameters for Cu$_n^-$ decay curves fitted to Eq. 
\ref{twocomp}. The power law exponent $p$, is equal to $-1+\delta$. 
The magnitude of the shorter time constants indicate that they 
are due to vibrational transitions. Uncertainties from the fits are 
$\pm$ 0.05 for the $\delta$'s and between 10 and 20\% for the lifetimes.
Vertical Detachment Energies, VDEs, from ref. \cite{Leopold1987} 
are given in the right column.}
\label{tab_fitparams}
\end{table}
We see that the size dependence of the rate constants does 
not seem to be influenced in any significant way by the 
well-known shell structure and odd-even effects observed in 
abundances of copper cluster anions \cite{Katakuse1986}, and also 
does not seem to reflect the vertical detachment energies \cite{Leopold1987}
(last column in Table \ref{tab_fitparams}). The power is expected to
be less than -1 for small clusters. 
In the simplest case, when $g(E)$ in Eq.\ref{eq_rateint} is constant 
and the heat capacity is not too small, the exponent $p$ is close 
to -1. Including the effect of the finite heat capacity
gives \cite{hansen01,andersen03,book}
\begin{eqnarray}
\label{delta}
p &\equiv& -1 + \delta \approx\\
\nonumber
&-&1 -\frac{1}{C} - \frac{2}{C} \cdot \frac{{\e}^{-\ln(\omega t)/C}}
{1-{\e}^{-\ln(\omega t)/C}}
\end{eqnarray}
in which $C$ is the effective microcanonical heat capacity in units of $k_B$
and $\omega$ the frequency factor in the rate constant of this channel. 

Although the short time $\delta$'s are negative as expected, the quantitative
agreement with Eq. \ref{delta} is poor. This is even more pronounced for
the slow component, where the value is large and positive for Cu$_6^-$.
This suggests that the values may be influenced by the excitation
energy distributions produced in the source. The idea can be further 
investigated by comparing the value of $\delta$ measured in photo-excitation 
experiments as the absorption of a photon and the resulting
enhanced decay probes the cluster's energy distribution at an 
excitation energy which is smaller than the energy at which the 
spontaneous decay occurs.
 
\begin{figure}[h]
\centering
\includegraphics[width=0.35\textwidth,angle=0]{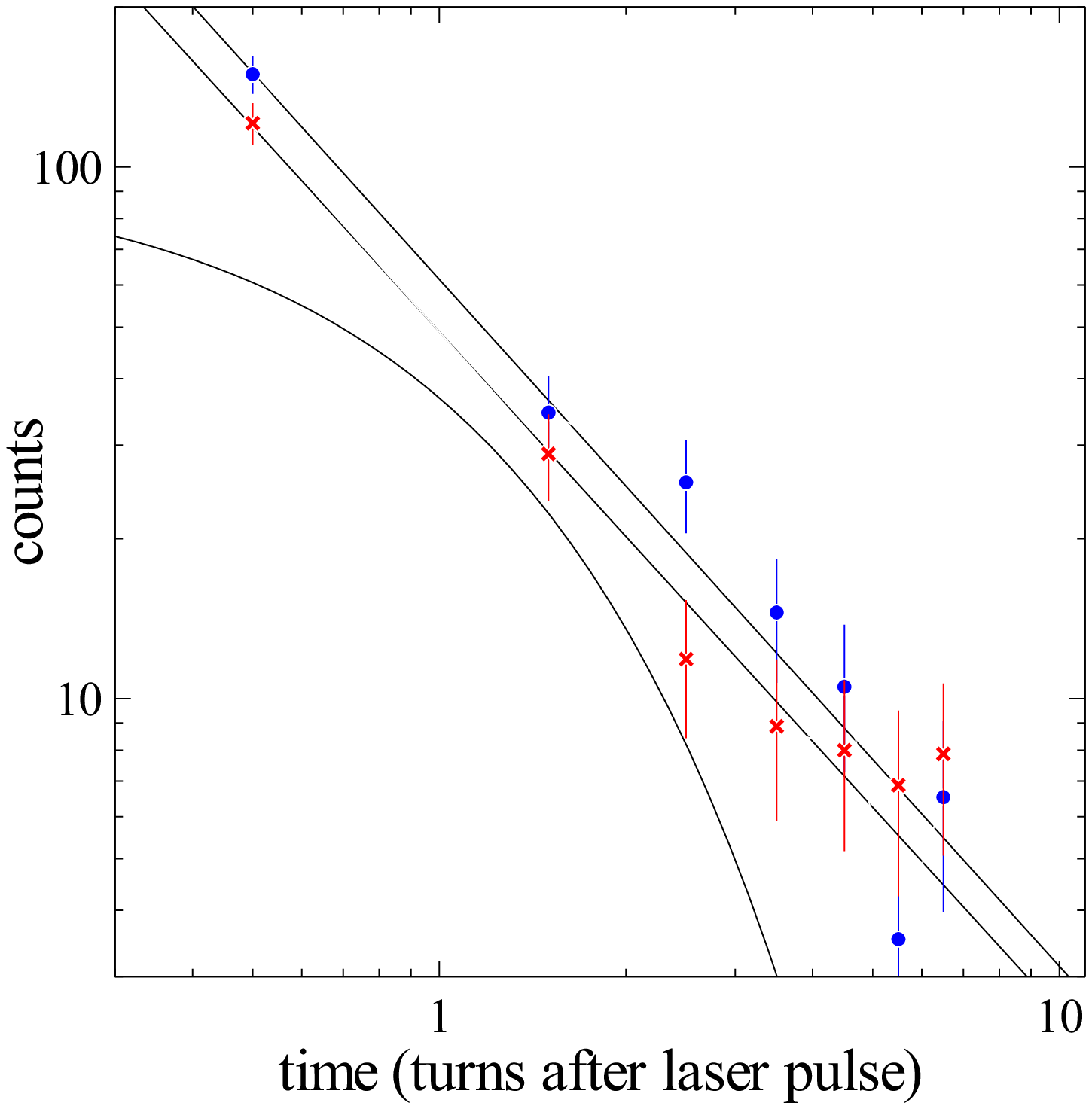}
\caption{A double-logarithmic plot of the photo-induced decay
of Cu$_6^-$ at the two wavelengths 850 nm (blue circles) and 
1100 nm (red crosses). The time unit is the revolution period 
for the Cu$_6^-$ ions in the ring which is $\sim$ 120 $\mu$s.
The lines are straight line fits. For comparison, an exponential 
decay with the lifetime of one period is plotted.\label{figCu6Laser}}
\end{figure}

The rates of decay of the signal induced by absorption of photons at
the wavelengths 850 nm and 1100 nm in Cu$_6^-$ are shown in 
Fig. \ref{figCu6Laser}. These two measurements show power-law decays 
with exponents that are very similar; -1.29 $\pm$ 0.08 for the 850 nm 
measurement and -1.28 $\pm 0.07$ for the 1100 nm measurement 
yielding a weighted average of -1.28 $\pm$ 0.05. 
This is identical to the value of $\delta$ in Table 
\ref{tab_fitparams} from the spontaneous decay at short times.
If the energy of the clusters that decay spontaneously is denoted 
by $E_0$, the distributions are thus probed at the three energies
$E_0$, $E_0 - 1.13$ eV and $E_0 - 1.45$ eV, yielding similar
$\delta$ values. Given that $E_0$ is close to the
threshold, these three points span a significant part of the
distribution. We thus conclude that a value of $\delta$ of -0.28 
reflects a property of the clusters themselves and not a property 
of the production process (i.e. not of the distribution $g(E)$).

It should be noted that laser excitation experiments, also on 
Cu$_6^-$, with photon energies 
of 1.165 eV (1064 nm) gave a time dependent exponent starting at 
$\sim -1.2$ at short 
times and reaching $-0.8$ at 2 s \cite{BreitenfeldtPRA2016}. 
The change is ascribed to 
radiative cooling and the related change of the excitation 
energy distribution. The energy distribution was found to agree
with a canonical distribution with an initial 
temperature of 1100 K (for times between 0.02 and 0.3 s). 
The canonical energy distribution at this
temperature changes slowly with energy at the energies probed 
in the experiments. Consequently, the $\delta$ value measured for these
times and photon energies should be essentially free from corrections
from the energy distribution and reflect intrinsic cluster properties.
This is consistent with the conclusions made above.

The present value, $\delta=-0.28$, for Cu$_6^-$ is, however, 
and as already mentioned,
not consistent with Eq. \ref{delta}. The equation gives -0.09 
and a direct simulation for Cu$_6^-$, analogous to that 
for Cu$_3^-$ gives values between -0.10 and -0.11 for electron 
detachment and atomic evaporation. We conclude 
that the measured value of $\delta$ is influenced by (as yet 
undetermined) factors intrinsic to the cluster, beyond the 
properties of the model used here. 
It is not clear to which degree this conclusion holds for the 
other cluster sizes.

\section{Summary}
In summary, we have measured the spontaneous decay of small copper 
cluster anions Cu$_n^-$ with $n=3-6$. We observe a complex decay 
behavior with two power-law decays each with their characteristic 
radiative lifetimes for Cu$_4^-$, Cu$_5^-$, and Cu$_6^-$. We do 
not observe radiative cooling for Cu$_3^-$. We tentatively 
identify the two populations for $n=4-6$ with cluster anions 
with lower and higher rotational excitations. For the higher rotational 
energies, elongated conformers have the lowest total energies for 
$n=4,5,6$, while lower rotational energies tend to
give more compact structures. The borders between the regions appear
to lie at angular momenta of a few hundreds units of $\hbar$, 
corresponding to rotational energies of a few tenths of an eV.
The trimer anion does not exhibit the two component decay seen for
the tetra-, penta- and hexamer. This is reasonable because here the
linear form of the anion has the lowest 
total energy for all values of the angular momentum quantum number.
Remarkably, no electronic cooling was observed 
in any of the measured clusters, which would have been manifested 
as sub-ms cooling times. The deviation of the initial power law decay 
for $n=4-6$ from -1 is only partly explained by the small heat 
capacity and seem to have contributions from unknown factors intrinsic 
to the clusters as indicated by the probing of distinctly different 
parts of the initial excitation energy distribution for Cu$_6^-$ 
by means of 
laser excitation.

As to the origin of the two angular momentum distributions for $n=4-6$, 
that we tentatively have identified with the two-component decay behaviors, 
we note the following. The radiative cooling time for each 
component depends on the geometry of the stored cluster anions. Provided 
the channel that dominates the signal at short times also has the shortest 
radiative decay time, this is sufficient to produce a two-component decay. 
Minority components with short decay times will remain unobserved. 
This classification refers to anion properties. 
The values of the $\delta$'s depend also on the precise decay 
channel, in the first instance through the number of vibrational degrees 
of freedom in the product. Different geometries may have different values
for that number, and the curves may therefore also depend on the crossing of 
energy curves in the upper part of the frames in Fig.\ref{skematisk}.
 
\section{Acknowledgements}
This work was supported by the Swedish Research Council (Contracts No. 
821-2013-1642, No. 621-2015-04990, No. 621-2014-4501, No. 621-2013-4084, 
and No. 2016-06625) and by the Knut and Alice Wallenberg Foundation. We 
acknowledge support from the COST Action No. CM1204 XUV/X-ray light and 
fast ions for ultrafast chemistry (XLIC). M.K. acknowledges financial 
support from the Mobility Plus Program (Project No. 1302/MOB/IV/2015/0) 
funded by the Polish Ministry of Science and Higher Education. 
\bibliographystyle{bibgen}

\bibliography{Cu3456}

\begin{thebibliography}{10}

\bibitem{Reinhed2009}
P.~Reinhed, A.~Orb\'an, J.~Werner, S.~Ros\'en, R.~D. Thomas, I.~Kashperka,
  H.~A.~B. Johansson, D.~Misra, L.~Br\"annholm, M.~Bj\"orkhage, H.~Cederquist,
  and H.~T. Schmidt.
\newblock {\em Phys. Rev. Lett.\/}, {\bf 103} (2009) 213002

\bibitem{Reinhed2010}
P.~Reinhed, A.~Orb{\'a}n, S.~Ros{\'e}n, R.~D. Thomas, I.~Kashperka, H.~A.~B.
  Johansson, D.~Misra, A.~Fardi, L.~{Br{\"a}nnholm}, M.~Bj{\"o}rkhage,
  H.~H.~Cederquist, and H.~T. Schmidt.
\newblock {\em Nuclear Instruments and Methods in Physics Research Section A:
  Accelerators, Spectrometers, Detectors and Associated Equipment\/}, {\bf 621}
  (2010) 83

\bibitem{Lange2010}
M.~Lange, M.~Froese, S.~Menk, J.~Varju, R.~Bastert, K.~Blaum, J.~R. Crespo
  López-Urrutia, F.~Fellenberger, M.~Grieser, R.~von Hahn, O.~Heber, K.~U.
  K{\"u}hnel, F.~Laux, D.~Orlov, M.~L. Rappaport, R.~Repnow, C.~D.
  Schr{\"o}ter, D.~Schwalm, A.~Shornikov, T.~Sieber, Y.~Toker, J.~Ullrich,
  A.~Wolf, and D.~Zajfman.
\newblock {\em Rev. Sci. Instrum.\/}, {\bf 81} (2010) 055105

\bibitem{thomas11}
R.~D. Thomas, H.~T. Schmidt, G.~Andler, M.~Bj{\"o}rkhage, M.~Blom,
  L.~Br{\"a}nnholm, E.~B{\"a}ckstr{\"o}m, H.~Danared, S.~Das, N.~Haag,
  P.~Halld{\'e}n, F.~Hellberg, A.~I.~S. Holm, H.~A.~B. Johansson,
  A.~K{\"a}llberg, G.~K{\"a}llersj{\"o}, M.~Larsson, S.~Leontein, L.~Liljeby,
  P.~L{\"o}fgren, B.~Malm, S.~Mannervik, M.~Masuda, D.~Misra, A.~Orb{\'a}n,
  A.~Pa{\'a}l, P.~Reinhed, K.~G. Rensfelt, S.~Ros{\'e}n, K.~Schmidt, F.~Seitz,
  A.~Simonsson, J.~Weimer, H.~Zettergren, and H.~Cederquist.
\newblock {\em Rev. Sci. Instrum.\/}, {\bf 82} (2011) 065112

\bibitem{vonHahn2011}
R.~von Hahn, F.~Berg, K.~Blaum, J.~R. Crespo Lopez-Urrutia, F.~Fellenberger,
  M.~Froese, M.~Grieser, C.~Krantz, K.~U. K{\"u}hnel, M.~Lange, S.~Menk,
  F.~Laux, D.~Orlov, R.~Repnow, C.~D. Schr{\"o}ter, S.~Shornikov, T.~Sieber,
  J.~Ullrich, A.~Wolf, M.~Rappaport, and D.~Zajfman.
\newblock {\em Nucl. Instrum. Methods Phys. Res., Sect. B\/}, {\bf 269} (2011)
  2871

\bibitem{Nakano2012}
Y.~Nakano, W.~Morimoto, T.~Majima, J.~Matsumoto, H.~Tanuma, H.~Shiromaru, and
  T.~Azuma.
\newblock {\em J. Phys.: Conf. Ser.\/}, {\bf 388} (2012) 142027

\bibitem{schmidt13}
H.~T. Schmidt, R.~D. Thomas, M.~Gatchell, S.~{Ros{\'e}n}, P.~Reinhed,
  P.~{L{\"o}fgren}, L.~{Br{\"a}nnholm}, M.~Blom, M.~Bj{\"o}rkhage,
  E.~B{\"a}ckstr{\"o}m, J.~D. Alexander, S.~Leontein, D.~Hanstorp,
  H.~Zettergren, L.~Liljeby, A.~K{\"a}llberg, A.~Simonsson, F.~Hellberg,
  S.~Mannervik, M.~Larsson, W.~D. Geppert, K.~G. Rensfelt, H.~Danared,
  A.~Pa{\'a}l, M.~Masuda, P.~Halld{\'e}n, G.~Andler, M.~H. Stockett, T.~Chen,
  G.~{K{\"a}llersj{\"o}}, J.~Weimer, K.~Hansen, H.~Hartman, and H.~Cederquist.
\newblock {\em Rev. Sci. Instrum.\/}, {\bf 84} (2013) 055115

\bibitem{andersen96}
J.~U. Andersen, C.~Brink, P.~Hvelplund, M.~O. Larsson, B.~Bech~Nielsen, and
  H.~Shen.
\newblock {\em Phys. Rev. Lett.\/}, {\bf 77} (1996) 3991

\bibitem{martin13}
S.~Martin, J.~Bernard, R.~{Br{\'e}dy}, B.~Concina, C.~Joblin, M.~Ji, C.~Ortega,
  and L.~Chen.
\newblock {\em Phys. Rev. Lett.\/}, {\bf 110} (2013) 063003

\bibitem{bernard08}
J.~Bernard, G.~Montagne, R.~Br{\'e}dy, B.~Terpend-Ordacière, A.~Bourgey,
  M.~Kerleroux, L.~Chen, H.~T. Schmidt, H.~Cederquist, and S.~Martin.
\newblock {\em Rev. Sci. Instrum.\/}, {\bf 79} (2008) 075109

\bibitem{ito14}
G.~Ito, T.~Furukawa, H.~Tanuma, J.~Matsumoto, H.~Shiromaru, T.~Majima, M.~Goto,
  T.~Azuma, and K.~Hansen.
\newblock {\em Phys. Rev. Lett.\/}, {\bf 112} (2014) 183001

\bibitem{goto13}
M.~Goto, A.~E.~K. {Sund{\'e}n}, H.~Shiromaru, J.~Matsumoto, H.~Tanuma,
  T.~Azuma, and K.~Hansen.
\newblock {\em J. Chem. Phys.\/}, {\bf 139} (2013) 054306

\bibitem{Najafian2014}
K.~Najafian, M.~S. Pettersson, B.~Dynefors, H.~Shiromaru, J.~Matsumoto,
  H.~Tanuma, T.~Furukawa, T.~Azuma, and K.~Hansen.
\newblock {\em J. Chem. Phys.\/}, {\bf 140} (2014) 104311

\bibitem{Kono2015}
N.~Kono, T.~Furukawa, H.~Tanuma, J.~Matsumoto, H.~Shiromaru, T.~Azuma,
  K.~Najafian, M.~S. Pettersson, B.~Dynefors, and K.~Hansen.
\newblock {\em Phys. Chem. Chem. Phys.\/}, {\bf 17} (2015) 24732

\bibitem{Ebara2016}
Y.~Ebara, T.~Furukawa, J.~Matsumoto, H.~Tanuma, T.~Azuma, H.~Shiromaru, and
  K.~Hansen.
\newblock {\em Phys. Rev. Lett.\/}, {\bf 117} (2016) 133004

\bibitem{froese11}
M.~W. Froese, K.~Blaum, F.~Fellenberger, M.~Grieser, M.~Lange, F.~Laux,
  S.~Menk, D.~A. Orlov, R.~Repnow, T.~Sieber, Y.~Toker, R.~von Hahn, and
  A.~Wolf.
\newblock {\em Phys. Rev. A\/}, {\bf 83} (2011) 023202

\bibitem{menk14}
S.~Menk, S.~Das, K.~Blaum, M.~W. Froese, M.~Lange, M.~Mukherjee, R.~Repnow,
  D.~Schwalm, R.~von Hahn, and A.~Wolf.
\newblock {\em Phys. Rev. A\/}, {\bf 89} (2014) 022502

\bibitem{BreitenfeldtPRA2016}
C.~Breitenfeldt, K.~Blaum, M.~W. Froese, S.~George, G.~Guzm{\'a}n-Ram{\'i}rez,
  M.~Lange, S.~Menk, L.~Schweikhard, and A.~Wolf.
\newblock {\em Phys. Rev. A\/}, {\bf 94} (2016) 033407

\bibitem{BackstromPRL2015}
E.~B\"ackstr\"om, D.~Hanstorp, O.~M. Hole, M.~Kaminska, R.~F. Nascimento,
  M.~Blom, M.~Bj\"orkhage, A.~K\"allberg, P.~L\"ofgren, P.~Reinhed, S.~Ros\'en,
  A.~Simonsson, R.~D. Thomas, S.~Mannervik, H.~T. Schmidt, and H.~Cederquist.
\newblock {\em Phys. Rev. Lett.\/}, {\bf 114} (2015) 143003

\bibitem{hansen01}
K.~Hansen, J.~U. Andersen, P.~Hvelplund, S.~P. {M{\o}ller}, U.~V. Pedersen, and
  V.~V. Petrunin.
\newblock {\em Phys. Rev. Lett.\/}, {\bf 87} (2001) 123401

\bibitem{Weisskopf}
V.~Weisskopf.
\newblock {\em Phys. Rev.\/}, {\bf 52} (1937) 295

\bibitem{book}
K.~Hansen.
\newblock {\em Statistical Physics of Nanoparticles in the Gas Phase\/},
  volume~73 of {\em Springer Series on Atomic, Optical, and Plasma Physics\/}.
\newblock Springer, Dordrecht (2013).
\newblock ISBN 978-94-007-5839-1

\bibitem{BeyerSwinehart}
T.~Beyer and D.~F. Swinehart.
\newblock {\em Commun. ACM\/}, {\bf 16} (1973) 379

\bibitem{Leopold1987}
D.~G. Leopold, J.~Ho, and W.~C. Lineberger.
\newblock {\em J. Chem. Phys.\/}, {\bf 86} (1987) 1715

\bibitem{Fedor05}
J.~Fedor, K.~Hansen, J.~U. Andersen, and P.~Hvelplund.
\newblock {\em Phys. Rev. Lett.\/}, {\bf 94} (2005) 113201

\bibitem{AvivPRA2011}
O.~Aviv, Y.~Toker, D.~Strasser, M.~L. Rappaport, O.~Heber, D.~Schwalm, and
  D.~Zajfman.
\newblock {\em Phys. Rev. A\/}, {\bf 83} (2011) 023201

\bibitem{KaflePRA2015}
B.~Kafle, O.~Aviv, V.~Chandrasekaran, O.~Heber, M.~L. Rappaport, H.~Rubinstein,
  D.~Schwalm, D.~Strasser, and D.~Zajfman.
\newblock {\em Phys. Rev. A\/}, {\bf 92} (2015) 052503

\bibitem{andersen03}
J.~U. Andersen, H.~Cederquist, J.~S. Forster, B.~A. Huber, P.~Hvelplund,
  J.~Jensen, B.~Liu, B.~Manil, L.~Maunoury, S.~{Br{\o}ndsted Nielsen}, U.~V.
  Pedersen, H.~T. Schmidt, S.~Tomita, and H.~Zettergren.
\newblock {\em Euro. Phys. J. D\/}, {\bf 25} (2003) 139

\bibitem{DebyeT}
C.~Ho, R.~Powell, and P.~Liley.
\newblock {\em J. Phys. Chem. Ref. Data\/}, {\bf 3, supplement 1} (1974) I–1

\bibitem{Andersen2002}
J.~U. Andersen, E.~Bonderup, and K.~Hansen.
\newblock {\em J. Phys. B: At. Mol. Opt. Phys.\/}, {\bf 35} (2002) R1

\bibitem{Katakuse1986}
I.~Katakuse, T.~Ichihara, Y.~Fujita, T.~Matsuo, T.~Sakurai, and H.~Matsuda.
\newblock {\em Int. J. Mass Spectrom. Ion Proc.\/}, {\bf 74} (1986) 33

\end{thebibliography}

\end{document}